\documentclass{llncs}

\usepackage{natbib}
\usepackage{graphicx}

\usepackage{booktabs}
\usepackage{tabu}
\usepackage{longtable}
\usepackage{multirow} 
\usepackage{tabularx}
\usepackage{lscape}
\usepackage{fancyhdr}
\usepackage{lastpage}

\usepackage{units}
 
\begin{document}

\title{A short review and primer on electroencephalography in human computer interaction applications}
\author{Lauri Ahonen\inst{1} and Benjamin Cowley\inst{1,2}}
\institute{Quantitative Employee unit, Finnish Institute of Occupational Health,\\
\email{benjamin.cowley@ttl.fi},\\
POBox 40, Helsinki, 00250, Finland
\and
Cognitive Brain Research Unit, Institute of Behavioural Sciences, University of Helsinki, Helsinki, Finland}

\maketitle              

\begin{abstract}
The application of psychophysiology in human-computer interaction is a growing field with significant potential for future smart personalised systems. Working in this emerging field requires comprehension of an array of physiological signals and analysis techniques. 

Methods to study central nervous system (CNS) are usually expensive and laborious. However, electroencephalography (EEG) is one of the most affordable and ambulatory methodologies for CNS research. It is in use in various clinical studies and have been broadly studied over decades. Despite that the recorded EEG signals are quite prone to noise and environmental factors it is the most widely used method in study of brain-computer interaction (BCI). Here we discuss briefly on various aspects of the recorded signals, their interpretation, and usage in the field of interaction studies.

This paper aims to serve as a primer for the novice, enabling rapid familiarisation with the latest core concepts. We put special emphasis on everyday human-computer interface applications to distinguish from the more common clinical or sports uses of psychophysiology.

This paper is an extract from a comprehensive review of the entire field of ambulatory psychophysiology, including 12 similar chapters, plus application guidelines and systematic review. Thus any citation should be made using the following reference:

{\parshape 1 2cm \dimexpr\linewidth-1cm\relax
B. Cowley, M. Filetti, K. Lukander, J. Torniainen, A. Henelius, L. Ahonen, O. Barral, I. Kosunen, T. Valtonen, M. Huotilainen, N. Ravaja, G. Jacucci. \textit{The Psychophysiology Primer: a guide to methods and a broad review with a focus on human-computer interaction}. Foundations and Trends in Human-Computer Interaction, vol. 9, no. 3-4, pp. 150--307, 2016.
\par}


\end{abstract}

\section{Background}

Recording electrical brain activity is useful in studies of human cognition. One of the techniques available, electroencephalography (EEG), provides this measurement in an affordable and non-invasive way. In addition, when compared to other functional brain activation recording techniques, it is relatively easy to set up, mobile, and suitable for recording outside a laboratory setting. The technique measures the summated activity in local populations of oriented neurons. This synchronous activation often elicits rhythmic oscillations with distinct frequencies. The subsequent volume conduction of these oscillations can be detected at the scalp. In the strictest terms, a neural oscillation is defined as an increase -- i.e., a spectral peak -- in the power of the EEG signal in a specific frequency band \citep{lopesdasilva_eeg_2013}. Therefore, although the word `oscillation' is used rather loosely in the context of EEG, a peak in the power spectrum should be present when one is seeking to identify an oscillation. A mere increase in power over a wider frequency domain does not constitute an oscillation in the frequency band of interest, however; on account of the weakness of neural currents, environmental electrical fields can cause significant interference in the signals measured, even for frequencies with detectable peaks. This section of the paper presents the possibilities that EEG affords for assessing cognitive and affective states in individuals. Here, we consider continuous EEG signals. Continuous signals encompass all EEG measurements of interest for our purposes apart from the event-related potentials (ERPs) that are discussed in \citet{cowley_primer_2016}


\begin{landscape}
\thispagestyle{empty}

\begin{longtable}{ccl}

\caption{One explanation for the functions in which various oscillatory systems are involved}
\label{tab.bandprop}\\

Band & Origin & Properties \\
\midrule
    \multirow{3}{*}{$\delta$} & \multirow{3}{*}{Thalamus} & In the study of event-related potentials (see Article ~\citep{cowley_primer_2016} for details), the P300 is a delta- and \\&& theta-band-related component \citep{basar_gamma_2001}. P300s are linked to attention and focus \\&& \citep{van_dinteren_2014}. Delta waves are seen mainly in deep sleep \citep{knyazev_eeg_2012}. \\

\cmidrule(lr){3-3}

\multirow{2}{*}{$\theta$} & \multirow{2}{*}{Hippocampus} & Frontal theta rhythms are considered to be `limbic fingerprints' on cortices \citep{lopes_da_silva_rhythmic_1992}. \\&& \citet{gruber_oscillatory_2006} suggests that there is familiarity-dependency in theta oscillations. \\ 

\cmidrule(lr){3-3}

\multirow{3}{*}{$\alpha$} &  \multirow{3}{*}{Thalamus} & Among the strongest oscillations seen in EEG signals, this carries information on working memory, \\&& alertness, and ability to focus \citep{noonan_distinct_2016}. It is characteristic of a relaxed but \\&& alert state of wakefulness. The $\alpha$ amplitude increases when the eyes are closed. \\ 

\cmidrule(lr){3-3}

    \multirow{2}{*}{$\mu$-rhythm} & \multirow{2}{*}{Parietal lobe} & Sensimotor rhythm, also known as $\mu$-rhythm, whose frequency range covers the $\alpha$ band and parts of \\&& the $\beta$ band. It is associated with maintaining or witnessing physical stillness \citep{tiihonen_magnetic_1989}. \\

\cmidrule(lr){3-3}

\multirow{3}{*}{$\beta$} & \multirow{3}{*}{Parietal lobe} & Low-frequency $\beta$ waves are often linked to busy thinking / anxiousness and active \\&& concentration. Furthermore, $\beta$ activity is related to the salience of a given stimulus \\&& and variance across emotionally charged tasks \citep{ray_eeg_1985}.\\

\cmidrule(lr){3-3}

$\gamma$ & Hippocampus & This band is linked to memory processes, problem-solving, fear, and consciousness \citep{merker_cortical_2013}. 
\end{longtable}
\end{landscape}

\section{Background}

The oscillatory analysis of an EEG signal consists of using spatial-, temporal-, and frequency-domain information for analysis of the volume conducted -- i.e., electrical currents outside neurons' axons (explained in more detail in the next paragraph). These currents are measured as voltage differences between electrodes distributed over the scalp. Work with the frequency domain was the historical foundation of EEG analysis. Ever since the discovery of alpha waves, by \citet{berger_uber_1929}, EEG has been considered a mixture of signals, with different frequencies, a perspective that has had strong implications for EEG analysis. Originally, four major types of sinusoidal EEG waves were recognised. These rhythms are presented, among other standard brain rhythms, in Table \ref{tab.bandprop}. In addition to temporal and frequency-domain features, one can analyse higher-order features such as synchrony and spatial distributions in EEG signals.

\begin{figure}[!t]
	\begin{center}
		\includegraphics[width=\textwidth]{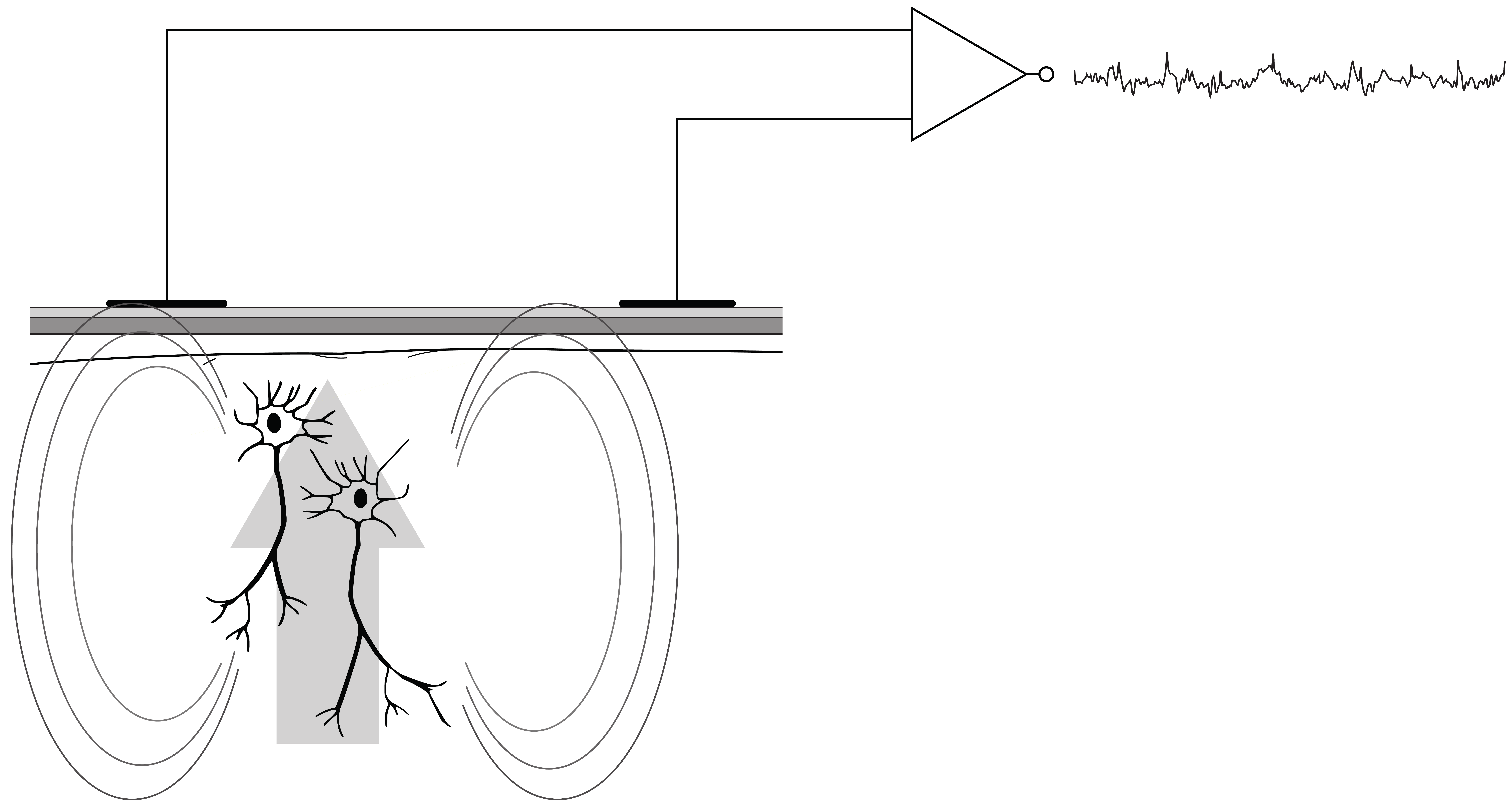}
	\end{center}
	\caption{Illustration of the volume conduction of neuronal firings recorded by EEG. Physically aligned and synchronously firing neuron populations (shown superimposed on the grey arrow) induce an electrical field (represented by concentric rings). This field propagates through the layers of dura, skull, and dermis (shown as horizontal lines) and can be measured at the scalp by means of distributed electrodes. Electrical fields that are strong enough to be measured are registered as a potential difference between distinct parts of the scalp, hence the need for a reference electrode.}
	\label{fig.eegalt}
\end{figure}

All non-invasive techniques employed for brain state evaluation require highly synchronised activity across neighbouring neuronal populations that results in signals measurable outside the scalp. The oscillations measurable on the surface of the scalp are rhythmic patterns caused by simultaneous pyramidal neuron action potentials in the area of interest. These patterns manifest as amplitude modulations via event-related synchronisation (ERS) or event-related desynchronisation (ERD) \citep{pfurtscheller_event-related_1999,guntekin_review_2014, horschig_hypothesis-driven_2014}. Figure~\ref{fig.eegalt} illustrates this phenomenon. In a concrete example, imagining a movement of the left hand leads to a contralateral ERD in the motor cortex (the right motor cortex for a movement of the left hand) in the $\mu$ and $\beta$ bands during imagining of the movement and to an ERS in the $\beta$ band (sometimes termed $\beta$ rebound) just after the imagined movement ends \citep{pfurtscheller_event-related_1999}.

One example of application is brain--computer interfaces (BCIs) \citep{vidal_toward_1973} utilising information provided by various frequency bands in EEG signals. A basic design for a motor-imagery BCI -- i.e., a set-up in which a computer is controlled via imagined movements -- would also exploit the spatial information by extracting features only from EEG channels localised over the motor areas of the brain, typically the channels C3 (lateral parietal) for the right hand's movements, Cz (medial parietal) for foot movements, and C4 (opposite lateral) for movements of the left hand. The frequency bands extracted in motor-imagery design are EEG-MU (referred to as `motoric $\mu$' herein) (\unit[8--12]{Hz}) and EEG-BETA (referred to as `$\beta$' herein) (\unit[16--24]{Hz}). The design focuses on temporal information in event-dependent amplitude variations in these features (spatial and spectral extractions). Some generalisations pertaining to frequency-domain information that are used in experiment designs of various types are listed in Table \ref{tab.bandprop}.

Finding a correlation between EEG features and mental states is far from a simple task, and caution should be employed in generalisation from the examples presented below. They are descriptions of results from single studies and previous reviews. Researchers utilise diverse techniques to extract features from the measured signals. In addition to the features introduced above, these techniques include statistical connectivity measures, complexity measures for evaluating the predictability of EEG signal features, and chaos-theory-inspired measures that consider signals' fractal dimensions. `Connectivity' denotes interplay of the spatially or spectrally separated signals. It is examined by assessment of basic correlation or, in more detail, by means of causality analysis (e.g., common spatial pattern (CSP) analysis), a direct transfer function (DTF), or similar techniques.

With ERD variations used as a feature, angry expressions elicit stronger beta and gamma oscillatory responses than happy and neutral expressions do when subjects view emotion-expressing faces \citep{guntekin_gender_2007, keil_effects_2001}. In addition, higher-power parietal beta activity has been found with emotionally loaded stimuli in comparison with non-emotional examples \citep{guntekin_event-related_2010}. The utility of simple power-based analysis techniques is limited and prone to session-to-session variations, inter-subject differences, and measurement inaccuracy. Analysing more than one EEG parameter improves the reliability of classification in respect of distinct brain functions and mental states. Results have been inconsistent between studies examining EEG band power changes elicited by emotional stimuli; hence, there is a need for in-depth analysis for all oscillatory responses, including phase-locked activity, evoked and/or induced power, time-frequency compositions, and connectivity measures. In addition, brain waves of various types are confounded by, for example, microsaccade potentials \citep{lopesdasilva_eeg_2013}.

According to \citet{guntekin_review_2014}, future studies should examine the synchronisation of alpha oscillatory responses when subjects are presented with an emotional stimulus, to reveal the dynamics of alpha activity in full. At the same time, it must be remembered that differences in age, gender, and phenotype greatly affect EEG dynamics, and that oscillatory systems are equally important in cognitive functions. For example, alpha oscillations in the occipital areas are directly linked to cognitive performance, showing a linear relationship with attentional processes in cognitive tasks. These findings suggest that human cognition is modulated by the phase of $\alpha$ oscillations \citep{ergenoglu_alpha_2004,de_graaf_alpha-band_2013}. Further studies support this claim, implying that even though this robust brain wave is linked to inhibitory processes, it does not represent an `idling' rhythm of the brain but, rather, expresses `windows of excitability' \citep{dugue_phase_2011}.

\section{Methods}

\paragraph{Instrumentation}
EEG devices take many forms, from consumer-grade one-channel `game controllers' to MRI-compatible medical instruments. These devices vary greatly, even at medical grade, in both their quality and properties such as number of channels, electrode type, and amplification techniques. For field experiments, EEG instrument development has produced a range of ambulatory research-grade devices. These are portable, utilise the latest innovations in electrode development, and measure a reasonable quantity of channels while remaining relatively easy to set up and use. At the time of writing, these include the 
LiveAmp system, from BrainProducts GmbH (Munich, Germany)\footnote{ See \url{http://pressrelease.brainproducts.com/liveamp/}.};
the Quasar DSI 10/20, from Quasar, Inc. (San Diego, CA, USA)\footnote{ See \url{http://www.quasarusa.com/products\_dsi.htm}.};
Enobio, by Neuroelectrics SLU (Barcelona, Spain)\footnote{ See \url{http://www.neuroelectrics.com/products/enobio/}.};
and g.MOBIlab+, from g.tec medical engineering GmbH (Schiedlberg, Austria)\footnote{ See \url{http://www.gtec.at/Products/Hardware-and-Accessories/g.MOBIlab-Specs-Features}.}.

\paragraph{Processing}
There are many ways to compute band power features from EEG signals \citep{herman_comparative_2008}. The canonical process flow for a basic temporal analysis of EEG data is as follows: Raw EEG~$\rightarrow$ Preprocessing~$\rightarrow$ Feature Extraction~$\rightarrow$ (Feature Selection)~$\rightarrow$ Classification (feature selection is an optional step used in subject-specified analysis).

A simple, popular, and efficient method is to band-pass filter the EEG signal from a given channel into the frequency band of interest, then square the resulting signal to compute the signal power and finally average it over time (e.g., over a time window of 1 s). 


Using more channels means extracting more features, which, in turn, increases the dimensionality of the data. High dimensionality leads to problems with classification and other mathematical analysis techniques. Accordingly, adding channels may even decrease performance if too few training data are available. Three main approaches can be applied in practice to exploit multiple EEG channels efficiently, all of which contribute to reducing dimensionality:

\begin{itemize}
\item \textbf{Feature selection algorithms} automatically select a subset of relevant features from among all the features extracted. These algorithms utilise machine learning techniques to perform the selection.
\item \textbf{Channel selection algorithms} are similar to feature selection methods. They utilise mathematical routines for automatic selection of a subset of relevant channels from all available channels.
\item \textbf{Spatial filtering algorithms} combine several channels into a single one, generally using weighted linear combinations. Features are extracted from the synthesised signal. 
Another kind of fixed spatial filter \citep{baillet_electromagnetic_2001} is represented by an inverse solution: an algorithm that enables one to estimate signals originating from sources within the brain on the basis of measurements taken at the scalp.
\end{itemize}

Alternatives for EEG feature representations can be divided into the following four categories: temporal, connectivity-, complexity-, and chaos-theory-related methods. Each class of methods extracts distinct attributes from EEG signals.

\paragraph{Temporal representations}

Temporal features quantify how signals vary over time. In contrast to the more basic features used for ERP, which consist simply of EEG amplitude samples over a short time window, some measures have been developed for characterising and quantifying variations in the signals measured. The corresponding features include Hjorth parameters \citep{hjorth_eeg_1970} and time-domain parameters (TDPs) \citep{vidaurre_time_2009}. Some research results have even suggested that TDPs could be more efficient than the gold-standard band power features \citep{vidaurre_time_2009}.

\paragraph{Connectivity measures}

Connectivity measures indicate how signals from two channels (or signals from two anatomical locations, obtained, for instance, via spatial filtering) are correlated and synchronised, or even whether one signal influences another one. In other words, connectivity features measure how the signals of two locations in spatial or spectral space are related. This is particularly useful for BCIs and mental state assessment, since it is known that, in the brain, there are long-distance communications between distant areas \citep{varela_brainweb_2001} and that the individual frequency bands are interconnected \citep{palva_phase_2005}. Therefore, connectivity features are put to increasing use in neuroscience and seem to be a very valuable complement to what are classed as more `traditional' features. Among connectivity features are coherence, phase-locking values, and the directed transfer function (DFT) \citep{varela_brainweb_2001,caramia_optimizing_2014,krusienski_value_2012}.

\paragraph{Complexity measures}

Complexity measures are used for ascertaining the complexity in EEG signals. The class of complexity measures quantify regularity or predictability of a signal. This has been shown to provide information about the cognitive state of a subject and, additionally, to provide information complementary to classical features such as band power features. Some of the features in this category are approximate entropy, a measurement adapted from anaesthesiology \citep{klockars_spectral_2006}; predictive complexity \citep{brodu_exploring_2012}; and waveform length \citep{lotte_new_2012}.

\paragraph{Chaos-theory-inspired measures}

The fourth category of features consists of chaos-related measures, used to estimate how chaotic an EEG signal is or which chaotic properties it might possess. Among the examples one could cite of corresponding features are the fractal dimension \citep{wang_real-time_2013} and multi-fractal cumulants \citep{brodu_exploring_2012}.

\section{Applications}

The complexity of human cognition does not map well to simplicity of analyses for low-spatial-accuracy signals such as EEG output, especially when the features extracted are rather simplified, as in the case of the five established frequency bands (EEG-DELTA, EEG-THETA, EEG-ALPHA, EEG-BETA, and EEG-GAMMA, referred to as $\delta,\,\theta,\,\alpha,\,\beta$, and $\gamma$). With such a small amount of information, it is obvious that direct mapping between cognitive processes and EEG oscillations is not realistic. That said, it is likely that EEG oscillations contribute to different cognitive functions, depending on their parameters (amplitude, frequency, phase, and coherence) and their spatial location within the brain. The related functions, together with corresponding references, are listed in Table \ref{tab.bandprop} and summarised below: 
\begin{itemize}
\item $\delta$: Attention and functional inhibition of executive functions \citep{harmony_functional_2013}
\item $\theta$: Hippocampal communications and the functional inhibition of executive functions \citep{colgin_mechanisms_2013}
\item $\alpha$: Pacemaker, timing and suppressing attention \citep{basar_review_2012}
\item $\beta$: Integrating aspects of motor and cognitive processes and affection \citep{kilavik_ups_2013}
\item $\gamma$: Conscious perception, updating of active memory, etc. \citep{merker_cortical_2013}
\end{itemize}

Besides the functional linkage -- i.e., the causality between distant cortex areas -- the interplay between frequency bands has received renewed attention. For instance, one hypothesis suggests that the capacity of working memory is represented by the ratio of brain activity involved at the theta and gamma frequencies \citep{lisman_storage_1995}. The nesting behaviour in different EEG oscillation bands itself was found long ago by \citet{berger_uber_1929}. More recently, the power of alpha waves has been demonstrated to modulate gamma oscillations, while it has been noted in addition that cross-frequency phase synchrony among $\alpha$, $\beta$, and $\gamma$ is related to cognitive functions \citep{palva_phase_2005}. It has been shown also that dysfunctions in these interplays are linked to disorders such as autism \citep{khan_local_2013}. For a review on cross-frequency couplings, see \citet{canolty_functional_2010} or \citet{herrmann_eeg_2015}.

In general, the information on cognitive functions conveyed by EEG signals has been found significant in more recent studies. For instance, the EEG traces of attention, motivation, and vigilance can be utilised in the context of learning through analysis of $\alpha$ power balance on the cortices. Besides the $\alpha$ frequency range, there is evidence of a $\delta$ power linkage on learning curves in `complex video-game' environments \citep{mathewson_different_2012}.

\section{Conclusion}

These more sophisticated analysis methods render it possible to gain information supplementary to, for instance, basic band power features, and they may increase classification accuracy in certain conditions.

The sophisticated methods presented here provide information that can be complementary to classical EEG analysis and thereby improve, for example, the classification performance of various machine learning algorithms \citep{lotte_tutorial_2014}.

The EEG signal processing methods presented here can be of great assistance in evaluating subject cognition, affection, and mental state. At this juncture, we must reiterate that classifying EEG signals is a tremendously complex task, on account of the non-stationary conditions, high dimensionality, artefacts, and the limited nature of the learning data. Current methods are far from perfect, but research is lively; for instance, methods are becoming more robust to noise, artefacts due to movement, session-to-session variations, etcetera. In turn, these advances are expected to lead towards more standardised analysis techniques for assessing human cognition. It is conceivable that they may even ultimately enable fully functional Brain-Computer Interfacing.

\bibliographystyle{plainnat}
\bibliography{ch5_eeg_bib}

\end{document}